\documentclass{article}
\usepackage{amsmath,amssymb}
\usepackage{xypic}
\usepackage{authblk}
\usepackage{color}
\usepackage{hyperref}
\usepackage[titletoc]{appendix}
\usepackage{fancyhdr}
\usepackage{lipsum}
\usepackage{MnSymbol}
\hypersetup{colorlinks=true,linkcolor=blue,citecolor=red}

\newtheorem{theorem}{Theorem}
\newtheorem{proposition}[theorem]{Proposition}
\newtheorem{lemma}[theorem]{Lemma}
\newtheorem{definition}[theorem]{Definition}

\newtheorem{remark}{Remark}

\newcommand{\proof}{\textbf{Proof: }}
\newcommand{\qed}{$\hfill\Box$}
\newcommand{\im}{\textrm{Im }}
\newcommand{\trqq}{T^*(\mathbb{R}\times Q\times Q)}
\newcommand{\trq}{T^*(\mathbb{R}\times Q)}

\author{Manuel de Le\'on \thanks{ mdeleon@icmat.es}}
\author{David Mart\'in de Diego\thanks{david.martin@icmat.es}}
\author{Miguel Vaquero \thanks{miguel.vaquero@icmat.es}}

\affil{Instituto de Ciencias Matem\'aticas,  ICMAT,\\
c$\backslash$ Nicol\'as Cabrera, nº 13-15, Campus Cantoblanco,UAM,\\
28049 Madrid, Spain}

\date{}
\begin{document}

\title{Hamilton-Jacobi theory, Symmetries and Coisotropic Reduction}
\maketitle
\begin{abstract} 
Reduction theory has played a major role in the study of
Hamiltonian systems. On the other hand, the Hamilton-Jacobi theory is one of
the main tools to integrate the dynamics of certain Hamiltonian
problems and a topic of research on its own. Moreover, the construction of several symplectic integrators rely on approximations of a complete solution of the Hamilton-Jacobi equation. The natural question that we address in this paper is
how these two topics (reduction and Hamilton-Jacobi theory) fit together. We obtain a reduction and
reconstruction procedure for the Hamilton-Jacobi equation with symmetries, even in a generalized
sense to be clarified below. Several applications and relations to
other reduction of the Hamilton-Jacobi theory are shown in the last
section of the paper. It is remarkable that as by-product we obtain a
generalization of the Ge-Marsden reduction procedure (\cite{Ge}) and
the results in \cite{GeEquivariant}. Quite surprinsingly, the classical
ansatzs available in the literature to solve the Hamilton-Jacobi
equation (see \cite{Ardema,Goldstein}) are also particular instances
of our framework.
\end{abstract}

\vfill
\newpage
\tableofcontents
%%%%%%%%%%%%%%%%%%%%%%%%%%%%%%%%%%%%%%%%%%%%%%%%%%%%%%%%%%%%%%%%%%%%%%%%%%INTRODUCTION
\section{Introduction}

The Hamilton-Jacobi theory is today a well-known theory by
mathematicians and physicist.  The equations
\begin{enumerate}
\item $\displaystyle H(q^i,\displaystyle\frac{\partial S}{\partial q^i}(q^i))=E$,

\item $\displaystyle \frac{\partial S}{\partial t}+H(t,q^i,\displaystyle\frac{\partial S}{\partial q^i}(t,q^i))=E$
\end{enumerate}
%\[
%\begin{array}{l}
%H(q^i,\displaystyle\frac{\partial S}{\partial q^i}(q^i))=E,
%\\ \noalign{\medskip}
%\displaystyle\frac{\partial S}{\partial t}+H(t,q^i,\displaystyle\frac{\partial S}{\partial q^i}(t,q^i))=E
%\end{array}
%\]
appear in any classical mechanics book, like \cite{AM,Goldstein}. The Hamilton-Jacobi theory is connected to geometric
optics and to classical and quantum mechanics in several intriguing ways. In geometric optics it
establishes the link between particles and waves through the {\it characteristic function}, \cite{holmI}. Hamilton and Jacobi extended this duality (wave-particle) to classical
mechanics, where a solution of the Hamilton-Jacobi equation allows the
reduction of the number of equations of motion by half, and a complete
solution of the Hamilton-Jacobi equation allows us to make a change of
variables that makes the integration of Hamilton's equations trivial (usually called ``a {\it transformation to equilibrium}''). A
detailed account of these topics can be found in
\cite{AM,Arnold}. Recently the Hamilton-Jacobi theory has also been
extended to the non-holonomic setting, \cite{cari,hjnoholonomo,iglesias,hamiltonjacobipoisson}.

The
Hamilton-Jacobi theory and the theory of generating functions also gave
rise to families of {\it symplectic numerical integrators} which over
long times are clearly superior to other methods (see
\cite{ChannellScovel,Kang,Ge}). Extending those integrators to the Lie-Poisson
setting motivated the beginning of the reduction of the
Hamilton-Jacobi theory in \cite{GeEquivariant,Ge}, by Z. Ge and
J.E. Marsden. After their approach, several works appeared along the
same lines,
\cite{BenzelGeScovel,ChannellScovelII,McLachlanScoveI,McLachlanScoveIIl,ScovelWeinstein}. Although
we are not dealing with numerical methods, getting a deeper
understanding of those results motivated this work to some
extent. Moreover, a general setting to develop numerical methods based
on the Hamilton-Jacobi theory for (integrable) Poisson manifolds will
appear elsewhere \cite{hamiltonjacobicompleto}. The importance of the
development of such {\it geometric-Poisson integrators} is beyond any doubt,
taking into account the success of their symplectic analogues.

On the other hand, reduction theory is still nowadays an important topic of
research. Since Jacobi's elimination of the node, and its
formalization through the Meyer-Marsden-Weinstein reduction, the
usefulness of the theory is widely known. A complete reference for
Hamiltonian reduction is \cite{hared}.

The present paper studies how to apply reduction theory to simplify
the Hamilton-Jacobi equation via the coisotropic reduction of
lagrangian submanifolds (see \cite{we}). We combine the aforementioned
coisotropic reduction and cotangent bundle reduction to obtain the
reduced Hamilton-Jacobi equation, which turns out to be an
algebraic-PDE equation. As mentioned above, previous attempts to obtain a reduction of
the (complete solutions of the) Hamilton-Jacobi equation were carried out by  Ge and Marsden  in \cite{Ge}
in order to provide a setting to develop Lie-Poisson
integrators. Nonetheless, they only work out the details in the case
where the configuration manifold is a Lie group, although they claim
that the general procedure can be obtained. The main difference
between their and our approach is that while Ge and Marsden reduce the generating function, say $S$, we focus on the corresponding lagrangian submanifold, say Im$(dS)$, that allows us to obtain a more general setting of wide applicability. For instance, generating functions which are not of {\it type I}, in the language of \cite{Goldstein} can be treated using our approach, while this seems not to be the case for the previous settings. Of course, Ge's framework can be obtained from our results in a
straightforward fashion as will be shown in the last section, where we
also deal with some examples, like a two particles Calogero-Moser
system. Finally, although we did not include it here, the results by
H. Wang in \cite{wang} are a particular case of our framework as
well. The use of generating families to obtain lagrangian submanifolds (\cite{Chaperon})
is another interesting topic not treated here that fits into our work.

The paper is organized as follows. In Section $2$ we provide the
necessary preliminaries and we establish the notation and conventions
that we follow during the rest of the paper. In Section $3$ we
introduce the Hamilton-Jacobi equation and the announced reduction and
reconstruction procedure. In Section $4$ we show that the ``two reduced
dynamics'' are related in the expected way. Section $5$ is devoted to
applications and examples. We also include two appendices about
adjoint bundles and magnetc terms to make the paper self-contained.

\section{Preliminaries}
In this paper all manifolds and mappings are supposed to be infinitely
differentiable ($C^{\infty}$). Given a map $f:M\rightarrow N$ between
manifolds $M$ and $N$, we will use the notation $Tf$ to denote the
tangent map  ($Tf:TM\rightarrow
TN)$, and $Tf(p)$, where $p$ is a point on $M$, to denote the tangent
map at that point $Tf(p):T_pM\rightarrow T_{f(p)}N$. Given a vector
field on the manifold $M$, say $X$, the evaluation of that vector field at a point $p\in M$
will read $X(p)$. The flow  of the vector fields under consideration will be assumed to be defined globally, although our results hold for locally defined flows with the obvious modifications. Along this paper $G$ will be a connected Lie group and $\mathfrak{g}$ the corresponding Lie algebra. We will make use  of $Ad^*$ to represent the Coadjoint action on the dual of $\mathfrak{g}$ given by
\[
\begin{array}{rccl}
Ad^*:&G\times \mathfrak{g}^*&\longrightarrow& \mathfrak{g}^* \\ \noalign{\medskip}
& (g,\mu)&\rightarrow & Ad^*_{g}(\mu)=\mu\circ TR_{g}\circ TL_{g^{-1}},
\end{array}
\]
where $L_g(h)=g\cdot h$ and $R_g(h)=h\cdot g$ are the left and right
multiplication on the group $G$. Notice that the Coadjoint action is a
left action. Given $\mu\in\mathfrak{g}^*$, $Orb^{Ad^*}(\mu)$ denotes the orbit by the Coadjoint action through $\mu$.

\subsection{Lifted actions to $TQ$ and $T^*Q$}

Let $G$ be a connected Lie group acting freely and properly on a manifold $Q$ by a left action $\Phi$
\[
\begin{array}{rccl}
\Phi:&G\times Q&\longrightarrow &Q \\\noalign{\medskip}
& (g,q) &\rightarrow & \Phi(g,p)=g\cdot p
\end{array}
\]
Given $g\in G$, we denote by $\Phi_g:Q \rightarrow Q$
the diffeomorphism defined by $\Phi_g(q)=\Phi(g,q)=g\cdot q$. Recall that under these conditions the quotient $Q/G$ can be endowed with a manifold structure such that the canonical projection $\pi:Q\rightarrow Q/G$ is a $G$-principal bundle. The action $\Phi$ introduced above can be lifted to actions on the tangent and cotangent bundles, $\Phi^T$ and $\Phi^{T^*}$  respectively. We briefly recall here their definitions.

\vspace{0.25cm}
$\bullet$\textbf{Lifted action on $TQ$}. We introduce the action $\Phi^T:G\times TQ\rightarrow TQ$
such that $\Phi_{g}^{T}:TQ\rightarrow TQ$ is defined by
\[
\Phi_g^{T}(v_q)=T\Phi_g(q)(v_q)\in T_{gq}Q
\quad 
\textrm{ for } v_q\in T_qQ.
\]

\vspace{0.25cm}
$\bullet$\textbf{Lifted action on $T^*Q$}. Analogously, we introduce the following  action $\Phi^{T^*}:G\times T^*Q\rightarrow T^*Q$ such that $\Phi_{g}^{T^*}:T^*Q\rightarrow T^*Q$ is defined by
\[
\Phi_g^{T^*}(\alpha_q)=(T\Phi_{g^{-1}})^*(gq)(\alpha_q)\in T_{gq}^*Q \quad\textrm{ for } \alpha_q\in T_q^*Q.
\]
 Both actions can be easily checked to be free and proper. If $\alpha_q\in T^*Q$, we will denote the orbit through $\alpha_q$ by
$Orb(\alpha_q)$.

%%%%%%%%%%%%%%%%%%%%%%%%%%%%%%%%%%%%%%%%%%%%%%%%%%%%%%%%%%%%%%%%%%%%%%%%%%%%%%%%%%%%%%%%
%%                         MOMENTUM MAPPING
%%%%%%%%%%%%%%%%%%%%%%%%%%%%%%%%%%%%%%%%%%%%%%%%%%%%%%%%%%%%%%%%%%%%%%%%%%%%%%%%%%%%%%%%
\subsection{Momentum Mapping}

As is well-known, there exists a $G$-e\-qui\-var\-iant momentum mapping for the
above action on $T^*Q$ with respect to its \textrm{canonical} symplectic form, from now on denoted by
$\omega_Q$. This momentum map is given by $J:T^*Q\rightarrow\mathfrak{g}^* $
where $J(\alpha_q)$ is such that
$J(\alpha_q)(\xi)=\alpha_q(\xi_{Q}(q))$ for $\xi\in\mathfrak{g}$. Here $\xi_Q$ is the vector field  on $Q$ determined via the action $\Phi$, called the infinitesimal generator. The integral curve of $\xi_Q$ passing through
  $q\in Q$ is just $t\rightarrow exp (t\xi)(q)$.

Given $\xi\in\mathfrak{g}$, we denote by $J_{\xi}:T^*Q\rightarrow
\mathbb{R}$ the real function obtained by the pairing between
$\mathfrak{g}$ and $\mathfrak{g}^*$, $J_{\xi}(\alpha_q)=\langle
J(\alpha_q),\xi\rangle$. By the definition of momentum mapping we have 
$\xi_{T^*Q}=X_{J_{\xi}}$,
where $\xi_{T^*Q}$ is the fundamental vector field generated by $\xi$
via the action $\Phi^{T^*}$. Indeed, we have
                             $i_{\xi_{T^*Q}}\omega_Q=dJ_{\xi}$ and $X_{J_\xi}$ is the vector field
  satisfying $i_{X_{J_\xi}}\omega_Q=d J_\xi$.

The next proposition, combined with the fact that $\Phi^{T^*}$ is free
and $G$ connected, ensures that for a connected Lie group every
$\mu\in \mathfrak{g}^*$ is a regular value and so $J^{-1}(\mu)$ is a
submanifold. In fact, the next proposition characterizes regular values of
momentum mappings taking into account the infinitesimal behavior of
the symmetries. We define $\mathfrak{g}_p=\{\xi\in\mathfrak{g}$
     such that $\xi_Q(p)=0\}$.

\begin{proposition}(Marsden {\it et al.} \cite{hared}) Let $(M,\
  \Omega)$ be a symplectic manifold and $G$ a Lie group which acts by
  symplectomorphism with equivariant momentum map $J$. An element
  $\mu\in\mathfrak{g}^*$ is a regular value of $J$ iff 
  $\mathfrak{g}_p =\{0\}$ for all $p\in J^{-1}(\mu)$.
\end{proposition}

\proof Let $p\in J^{-1}(\mu)$ and assume that $\mathfrak{g}_p=0$, then we will show that
$TJ(p)$ is surjective. This is equivalent to proving that the
anihilator of Im($TJ(p)$) is $\{0\}$. Assume that $\xi\in\mathfrak{g}$
is such that the natural pairing $\langle TJ(p)(X),\xi\rangle=0$ for all
for all $X\in T_pS$. That means that $TJ(p)(X)(\xi)=0$ or that
$\Omega(X(p),\xi_M(p))=0$ for all $X\in
T_pS$. Since $\Omega$ is non-degenerate that means $\xi_M(p)=0$ and so
$\xi\in \mathfrak{g}_p$ and therefore $\xi=0$ by hypothesis. Reversing the
computation the converse easily follows.

\qed

\begin{remark} {\rm In the case that concerns us, namely $(T^*Q, \ \omega_Q)$
  with the action $\Phi^{T^*}$, the previous theorem
  says that $J^{-1}(\mu)$ is always a submanifold of $T^*Q$.}
\end{remark}

Now we introduce $G$-invariant lagrangian submanifolds and the main
results about them. The main results of this paper will be direct applications of these results.
\begin{definition} Assume as above that the triple $(T^*Q,\omega_Q,h)$
  is endowed with a hamiltonian action $\Phi$. A $G$-invariant
  lagrangian submanifold is  a
lagrangian submanifold $L$ in $T^*Q$ such that for all $g\in G$ we have $\Phi^{T^*}_g(L)=L$.
\end{definition}

We give a characterization of $G$-invariant lagrangian submanifolds in
terms of equivariant momentum mappings. The next result should be
considered as a generalization of the Hamilton-Jacobi theory, an
explanation for this claim will be given in Remark \ref{remark2}. More
detailed results in this direction are given in
\cite{MarreroLagrangian}.
 \begin{lemma}\label{lemma} Under the previous assumptions, let $L\subset T^*Q$ be a  lagrangian submanifold of
 $(T^*Q, \ \omega_Q)$. Then $J$ is constant along $L$ if and only if $L$ is $G$-invariant. 
\end{lemma}
 \proof Let be $\alpha_q\in L$ and $X\in T_{\alpha_q}L$, then 
 \begin{equation}\label{one}
 dJ_{\xi}(\alpha_q)(X)=(i_{\xi_{T^*Q}}\omega_Q)(\alpha_q)(X)=\omega_Q(\alpha_q)(\xi_{T^*Q}(\alpha_q),X).
 \end{equation}
Now, notice that
\[
\xi_{T^*Q}(\alpha_q)=\textrm{tangent vector at $t=0$ to the curve $exp(t\xi)(\alpha_q)$.}
\]
Since $exp(t\xi)(\alpha_q)$ is contained in the orbit of $\alpha_q\in L$,
and $Orb(\alpha_q)\subset L$ since $L$ is $G$-invariant (that is, $G\cdot
L\subset L$), we deduce that $\xi_{T^*Q}(\alpha_q)\in T_{\alpha_q}L$.
Therefore, \eqref{one} vanishes since $L$ is lagrangian. Finally, since
$J_{\xi}$ is constant along $L$, we have $J_{\xi}(\alpha_q)=c_\xi$ for all
$\alpha_q\in L$ and for all $\xi\in\mathfrak{g}$ and thus,
$J(\alpha_q)=\mu$ for all $\alpha_q\in L$ (such that $\mu(\xi)=c_\xi$). Reversing the computations we obtain the other implication.

\qed
\vspace{0.5cm}

\begin{remark}\label{remark2} {\rm Notice that the Hamilton-Jacobi
    theory itself is a particular case of the theorem above, which
    should be considered as a generalization of that theory. Let us
    clarify this assertion, we have a hamiltonian system
    ($T^*Q,\omega_Q,h$), with the associated hamiltonian vector field
    $X_h\in \mathfrak{X}(T^*Q)$; we denote the flow of $X_h$ by
    $\Psi^h:\mathbb{R}\times T^*Q\rightarrow T^*Q$, recall that the
    flow is just an $\mathbb{R}$ action on $T^*Q$. By Liouville's
    Theorem this action is hamiltonian and it is easy to see that the
    hamiltonian $h$ is a momentum map for that action. If we seek a
    $\mathbb{R}$-invariant lagrangian submanifold, say $L$, then, by
    Lemma \ref{lemma} $h_{|L}=E$, where $E$ is a constant. Moreover, assume that $L=\textrm{Im}(dS)$ where $S:Q\rightarrow \mathbb{R}$ is a real function, then we recover the classical Hamilton-Jacobi equation \[H(q^i,\displaystyle\frac{\partial S}{\partial q^i}(q^i))=E.\] The time-dependent and complete solutions cases of the Hamilton-Jacobi equation follow by an analogous construction.}
\end{remark}

\begin{remark} {\rm Most of the results in \cite{GeEquivariant} can be
  recovered from  Lemma \ref{lemma}. Indeed, there the author claims that
  there is a deep connection between the symmetry of a symplectic
  difference scheme and the preservation of first integrals. For instance, in \cite{GeEquivariant}
  p. $378$, the following theorem is stated

\vspace{\topsep}
\noindent {\bf Theorem} {\it  A symplectic difference scheme preserves a function $f$
up to a constant
\[
f\circ D_h=f+c
\]
iff the scheme is invariant under the phase flow of $f$.
}
\vspace{\topsep}

Assume a hamiltonian system ($M,\Omega,H$). Then 
a symplectic scheme  (following \cite{GeEquivariant}, p. $377$),
after fixing Darboux coordinates on $M$, is a rule which assigns to every
hamiltonian function a symplectic map depending smoothly on a
paramenter $\tau$, called the time step. A symplectic difference
scheme is denoted in \cite{GeEquivariant} by $D_h^\tau$. This means that a
symplectic difference scheme is just a lagrangian submanifold graph$(D_h^\tau)=L\subset
M\times M$, with the symplectic structure
$\overline{\Omega}=\pi_1^*\Omega-\pi_2^*\Omega$ on $M\times M$, where $\pi_i:M\times
M\rightarrow M$ are the
corresponding projections over the $i$-factor. Consider now
$\overline{f}=\pi_1^*f-\pi_2^*f$, it is obvious that $f$ is
preserved by $D_h^\tau$ iff $\overline{f}$ is constant along
graph$(D_h^\tau)=L$. By definition $X_{\overline{f}}=(X_f,X_f)$ and a
straightforward application of Lemma \ref{lemma}, taking into account
that $\overline{f}$ is the moment of the action give by the flow of
$X_{\overline{f}}$, gives that $\overline{f}$ is constant along $L$
iff $L$ is invariant under the flow of
$X_{\overline{f}}$. Recalling that the first statement is equivalent
to $f$ being preserved by the symplectic scheme and the second claim
is equivalent to saying that the scheme is invariant under the phase
flow we recover the main ``principle'' of \cite{GeEquivariant}.}
\end{remark}

\begin{remark}{\rm
Assume that we are in the hypothesis of the above lemma. If
$J(L)=\{\mu\}, \ \mu\in\mathfrak{g}^*$, then we deduce that $\mu$ is a
fixed point for the Coadjoint action $Ad^*:G\rightarrow
Aut(\mathfrak{g}^*)$. Indeed, remember that $J$ is $G$-equivariant, that is, the following diagram is commutative
\[
\xymatrix{
T^*Q \ar[rr]^{J}&& \mathfrak{g}^* \\ 
\\
T^*Q \ar[rr]^{J} \ar[uu]^{\Phi^{T^*}_g}&&\mathfrak{g}^* \ar[uu]_{Ad^*_{g^{-1}}}
}
\]
Then $Ad^*_{g^{-1}}(\mu)=Ad^*_{g^{-1}}J(\alpha_q)=J( \Phi^{T^*}_g(\alpha_q))=\mu$, for all $g\in G$.}
\end{remark}

\begin{lemma}\label{25}
If $\mu$ is such that $G_{\mu}=G$, then $J^{-1}(\mu)$ is a coisotropic submanifold ($G_{\mu}$ denotes the isotropy group with respect to the
Coadjoint action).
\end{lemma}

\proof Given any point  $\alpha_q\in J^{-1}(\mu)$, we have

\[
\begin{array}{ll}
T_{\alpha_q}J^{-1}(\mu)&=\textrm{ker}(TJ({\alpha_q}))=\{X\in T_{\alpha_q}(T^*Q) \textrm{ such that } TJ({\alpha_q})(X)=0\}\\ \noalign{\medskip}&=\{X\in T_{\alpha_q}(T^*Q) \textrm{ such that } TJ({\alpha_q})(X)(\xi)=0 \textrm{ for all }\xi\in \mathfrak{g}\}\\\noalign{\medskip}&=\{X\in T_{\alpha_q}(T^*Q) \textrm{ such that } \omega({\alpha_q})(X,\xi_{T^*Q})=0 \textrm{ for all }\xi\in \mathfrak{g}\} \\ \noalign{\medskip}&=(T{\it Orb({\alpha_q})})^{\perp}
\end{array}
\]
since $\xi_{T^*Q}(\alpha_q)$ generates the orbit through $\alpha_q$. Therefore, we have
 $(T_{\alpha_q}J^{-1}(\mu))^{\perp}$ $=T_{\alpha_q}{\it Orb({\alpha_q})}$
for all $\alpha_q\in J^{-1}(\mu)$. But $J$ is $G$-equivariant and $G=G_\mu$, thus $
J(\Phi^{T^*}_g(\alpha_q))=Ad^*_{g^{-1}}J(\alpha_q)=Ad^*_{g^{-1}}\mu=\mu$  and so $
\Phi^{T^*}_g(\alpha_q)\in J^{-1}(\mu)$. Then, $Orb(\alpha_q)\subset J^{-1}(\mu)$, and thus $
T_{\alpha_q}Orb(\alpha_q)\subset T_{\alpha_q}J^{-1}(\mu)$.
Consequently, we have $
(T_{\alpha_q}J^{-1}(\mu))^{\perp}=T_{\alpha_q}Orb({\alpha_q})\subset
T_{\alpha_q}(J^{-1}(\mu))$ and we conclude that $ J^{-1}(\mu)$ is coisotropic.

\qed

The next result is a well-known theorem in symplectic geometry, see \cite{kost,we}. It will allow us to carry out our reduction procedure in a straightforward way.

\begin{theorem}[Coisotropic Reduction]\label{reductionlagrangian}
Let $(M,\omega)$ be a symplectic manifold,  $C\subset  M$ a
coisotropic submanifold and $C/\hspace{-1.5mm}\sim$ the quotient space of $C$ by the
characteristic distribution $D=\textrm{ker}(\omega_{|C})$; we shall
denote by $\pi:C\rightarrow C/\hspace{-1.5mm}\sim$ the canonical projection and by
$\omega_C$ the natural projection of $\omega$ to $C/\hspace{-1.5mm}\sim$ (notice that
$(C/\hspace{-1.5mm}\sim,\ \omega_C)$ is again a symplectic manifold,
assuming that it is again a manifold). Assume that $L\subset M$ is a lagrangian submanifold such that $L\cap
C$ has clean intersection, then $\pi(L\cap C)$ is a lagrangian
submanifold of $(C/\hspace{-1.5mm}\sim,\ \omega_C)$. 
\end{theorem}

The following diagram illustrates the above situation
\[
\xymatrix{
L\cap C \ar@{^{(}->}[rr]^{i_{L\cap C}}\ar[dd]^{\pi}&& C \ar[rr]^{i_{C}}\ar[dd]^{\pi}&& M \\ \\
\pi(L\cap C)\ar@{^{(}->}[rr]_{i_{\pi(L\cap C)}}&& C/\sim & 
}
\]

We can apply this theorem to the situation described before. Indeed, given
$\mu\in\mathfrak{g}^*$ such that it is a fixed point of the Coadjoint
action (i.e. $Ad^*_g(\mu)=\{\mu\}$ for all $g\in G$), then we have the
following diagram, since by Lemma \ref{25} we know that $J^{-1}(\mu)$ is coisotropic:
\[
\xymatrix{
 J^{-1}(\mu) \ar[rr]^{i_{J^{-1}(\mu)}}\ar[dd]^{\pi'}&& T^*Q \\ 
\\
 J^{-1}(\mu)/\textrm{ker}({\omega_Q}_{|J^{-1}(\mu)}) && 
}
\]

But  $\textrm{ker}({\omega_Q}_{|J^{-1}(\mu)})(\alpha_q)=(T_{\alpha_q}J^{-1}(\mu))^{\perp}=T_{\alpha_q}{\it
  Orb(\alpha_q)}$ for all $\alpha_q\in J^{-1}(\mu)$,  and since
$G=G_{\mu}$, we can see that
$J^{-1}(\mu)/\textrm{ker}({\omega_Q}_{|J^{-1}(\mu)})=J^{-1}(\mu)/G$. But this is just the symplectic reduction of $T^*Q$ according to the
Marsden-Weinstein reduction theorem, see \cite{MaWe}.

%%%%%%%%%%%%%%%%%%%%%%%%%%%%%%%%%%%%%%%%%%%%%%%%%%%%%%%%%%%%%%%%%%%%%%%%%%%%%
%%                 THE HAMILTON-JACOBI EQUATION
%%%%%%%%%%%%%%%%%%%%%%%%%%%%%%%%%%%%%%%%%%%%%%%%%%%%%%%%%%%%%%%%%%%%%%%%%%%%%

\section{The Hamilton-Jacobi Equation}
\subsection{Generalized Solutions}

Along this section $h:T^*Q\rightarrow \mathbb{R}$ will be a  hamiltonian function. We are going to use the previous results to carry out our reduction of the Hamilton-Jacobi equation. By Hamilton-Jacobi equation we mean
\begin{enumerate}
\item {\it The time-independent Hamilton-Jacobi equation:} \[\displaystyle h(q^i,\displaystyle\frac{\partial S}{\partial q^i}(q^i))=E.\]
\item {\it The time-dependent Hamilton-Jacobi equation:} \[\displaystyle \frac{\partial S}{\partial t}+h(t,q^i,\displaystyle\frac{\partial S}{\partial q^i}(t,q^i))=E.\]
\item {\it A complete solution of the Hamilton-Jacobi equation:} that is, a real-valued
  function $S(t,q^i,\alpha^i)$ depending on as many parameters $(\alpha^i)$ as the dimension of the configuration manifold, such that
\begin{enumerate}
\item For every  (fixed) value of the parameters $(\alpha^i)$,
  $S(t,q^i,\alpha^i)$ satisfies the time-dependent Hamilton-Jacobi equation,
\[
\displaystyle\frac{\partial S}{\partial t}(t,q^i,\alpha^i)+h(t,q^i,\displaystyle\frac{\partial S}{\partial q^i}(t,q^i,\alpha^i))=0.
\]
\item The {\it non-degeneracy condition}: consider the matrix whose components
  $(i,j)$  are given by $\displaystyle\frac{\partial^2 S}{\partial q^i\partial
    \alpha^j}$, that we denote by $(\displaystyle\frac{\partial^2 S}{\partial q^i\partial
    \alpha^j})$, then \[det(\displaystyle\frac{\partial^2 S}{\partial q^i\partial
    \alpha^j})\neq 0.\]
\end{enumerate}
\end{enumerate}

We define below the concept of generalized solution, which is a
generalization of a solution of the time-independent Hamilton-Jacobi
equation (see \cite{Cardin}), and we develop our theory for this case. Analogous procedures hold for the time-independent Hamilton-Jacobi equation and for the complete solutions cases, as both settings can  be (almost) considered as particular cases of the time-independent Hamilton-Jacobi theory. Along the examples section, sections \ref{time-dependentfortimeindependent} and \ref{completesolution}, we will make this claim explicit.

\begin{definition}
We say that a submanifold $L\subset T^*Q$ is a  solution of the (time-independent) Hamilton-Jacobi problem for $h$, if:
\begin{itemize}
\item $L$ is a lagrangian submanifold of $T^*Q$.
\item $h$ is constant along $L$.
\end{itemize}

A solution $L$ of the Hamilton-Jacobi equation for $h$ is {\it
  horizontal} if $L=$Im$(\gamma)$, being $\gamma$ a $1$-form on $Q$.
\end{definition}

\begin{remark}{\rm
Let us describe with more detail the case of horizontal solutions, that is, when
 $L=\textrm{Im}(\gamma)$, $\gamma$ a
1-form on $Q$. Recall that Im($\gamma$) is lagrangian if and only if $\gamma$ is
closed, so locally
\[
\gamma=dS.
\]
Therefore, the condition $h_{|\textrm{Im}(\gamma)}=cte$, can be
equivalently written as
\[
h\circ\gamma=cte
\]
or
\[
h(q^i,\frac{\partial S}{\partial{q^i}})=cte
\]
which is the usual form of the Hamilton-Jacobi equation. This fact justifies the definition above.}
\end{remark}
\begin{remark}{\rm
Notice that the fact that a horizontal lagrangian submanifold $L$ is
$G$-invariant does not imply that its generating function is invariant
too. In fact, its generating function will be invariant iff
$J(L)=0$. Since $J\circ dS=\mu$, then $dS(q)(\xi_Q(q))=J\circ dS(q)
(\xi_{T^*Q}(q))=\mu(\xi)$, which only vanish for all
$\xi\in\mathfrak{g}$ if $\mu=0$. Here the advantages of dealing with
lagrangian submanifolds instead of functions are already manifest, as
there are $G$-invariant lagrangian manifolds whose generating function
is not $G$-invariant, see Section \ref{examples}. Notice that
invariance of the generating function has been assumed in \cite{GeEquivariant,Ge}.}
\end{remark}
%%%%%%%%%%%%%%%%%%%%%%%%%%%%%%%%%%%%%%%%%%%%%%%%%%%%%%%%% Invariant G-solution
\subsection{Invariant $G$-solutions}

We assume now that a Lie group $G$ acts on $Q$ such that the action is
free and proper. Given $\mu\in\mathfrak{g}$, then $J^{-1}(\mu)$ is a
submanifold of $T^*Q$. We can summarize the situation in the following diagram:
\[
\xymatrix{
J^{-1}(\mu) \ar@{^{(}->}[rr] && T^*Q \ar[dd]^{\pi_Q}\ar[rr]^{J}\ar[ddrr]^{\pi} && \mathfrak{g}^* \\
\\
                          && Q \ar[rrdd]^{\pi_G}&& T^*Q/G \ar[dd]^{p}\\
\\
                          &&     && Q/G, 
}
\]
where $\pi_Q$, $\pi$, $\pi_G$ and $p$ are the canonical projections. We will use $\pi'$ for the projection
\[
\pi'=\pi_{|J^{-1}(\mu)}: J^{-1}(\mu)\rightarrow T^*Q/G.
\]

As we know, $T^*Q/G$ has a Poisson structure induced by the canonical symplectic structure on $T^*Q$,
such that.
\[
\pi:T^*Q \rightarrow T^*Q/G
\]
is a Poisson morphism (see \cite{hared} for the details). The next proposition shows the symplectic structure of the leaves of
the characteristic distribution of the Poisson structure of $T^*Q/G$.

\begin{proposition}(Marsden {\it et al.} \cite{kost,hared,sour})
The symplectic leaves of $T^*Q/G$ are just the quotient spaces $\left(J^{-1}({\it Orb}^{Ad^*}(\mu))\right)/G$.
\end{proposition}

\subsection{Reduction and Reconstruction}
Assume now that $\mu$ is a fixed point for the Coadjoint action, i.e. $Orb^{Ad^*}(\mu)=\{\mu\}$. Then  $J^{-1}(\mu)/G$
is a symplectic leaf of $T^*Q/G$. Assume now that $L\subset J^{-1}(\mu)$ is a lagrangian submanifold;
since $J^{-1}(\mu)$ is a coisotropic submanifold of $(T^*Q, \
\omega_Q)$, we deduce that $\pi(L)$ is a lagrangian submanifold of
the quotient $J^{-1}(\mu)/G$ by applying the Coisotropic Reduction
Theorem. Obviously, the condition of clean intersection is
trivially satisfied.

In reference \cite{hared} it is shown that $J^{-1}(\mu)/G$ is
diffeomorphic to the cotangent bundle $T^*(Q/G)$. Moreover,
considering the symplectic structure $\omega_{\mu}$ on $J^{-1}(\mu)/G$
given by the Marsden-Weinstein reduction procedure, the two manifolds
 are symplectomorphic, where on $T^*(Q/G)$ we are considering the
symplectic structure given by the canonical one plus a magnetic term
$\omega_{Q/G}+B_{\mu}$ see (Appendix \ref{B}). Combining the last two paragraphs we can
see $\pi(L)$ as a lagrangian submanifold of a cotangent bundle with a
modified symplectic structure. We proceed now to sketch the aforementioned
identification using a connection. Recall that $\pi_G:Q\rightarrow Q/G$ is a $G$-principal fiber bundle with the
structure group $G$. A connection $A$ on $\pi_G:Q\rightarrow Q/G$ induces a splitting
\begin{equation}\label{two}
T^*Q/G\equiv T^*(Q/G)\times_{Q/G}\tilde{\mathfrak{g}}^*
\end{equation}
(see \cite{hared} for a detailed discussion of this splitting) where $\tilde{\mathfrak{g}}^*$ denotes the adjoint bundle to
$\pi_Q:Q\rightarrow Q/G$ via the Coadjoint representation, $\tilde{\mathfrak{g}}^*=Q\times_G\mathfrak{g}^*$ (see \cite{hared} and Appendix A for a description of this bundle). The identification \eqref{two} is given by
\[
\begin{array}{rccl}
\Psi:& T^*Q/G& \longrightarrow &T^*(Q/G)\times_{Q/G}\tilde{\mathfrak{g}}^* \\ \noalign{\medskip}
&[\alpha_q]&\rightarrow& \Psi([\alpha_q])=[(\alpha_q\circ \bold{h},J(\alpha_q))],
\end{array}
\]
where $\bold{h}$ represents the horizontal lift
$T_{\pi_G(q)}(Q/G)\rightarrow T_qQ$ of the connection $A$. Therefore, we have
\[
\xymatrix{
T_{\pi_G(q)}(Q/G) \ar[rr]^{\bold{h}} \ar[drdr]_{\alpha_q\circ \bold{h}}&& T_qQ \ar[dd]^{\alpha_q}\\
\\
                                    &&\mathbb{R}.
}
\] 
If $\alpha_q\in J^{-1}(\mu)$ then $ J(\alpha_q)=\mu$, and
$\Psi([\alpha_q])=(\alpha_q\circ \bold{h}, J(\alpha_q)=\mu)$, so that
$J^{-1}(\mu)/G$ can be identified with $T^* (Q/G)$.
\[
\Psi(J^{-1}(\mu)/G)=T^*(Q/G)\times_{Q/G}(Q\times \{\mu\}/G)\equiv T^*(Q/G)
\]

\begin{remark}{\rm
Notice that $dim(Q)=n$, and then $\dim(J^{-1}(\mu))=2m-k$ where $\dim(G)=k$. Thus,
$\dim(J^{-1}(\mu)/G)=2n-k-k=2(n-k)$ and $\dim(T^*(Q/G))=2(n-k)$.}
\end{remark}

Notice that $J^{-1}(\mu)/G$ and $T^*(Q/G)$ are not only diffeomorphic, moreover, it is
possible to show that they are symplectomorphic, while $J^{-1}(\mu)/G$
is considered as a symplectic leaf of $T^*Q/G$ and $T^*(Q/G)$ is
equipped with the canonical symplectic structure modified by a magnetic
term (it is explained in the cited paper, \cite{hared}, and the magnetic term
$\beta_{\mu}$ comes from the connection $A$,
$\omega_{Q/G}+\beta_{\mu}$). If $\mu=0$, then the magnetic term vanishes and we have the
canonical symplectic structure $\omega_{Q/G}$.

Next, we consider a $G$-invariant hamiltonian $h$ on $T^*Q$. Then we
have
\[
\xymatrix{
T^*Q \ar[dd]^{\pi} \ar[rr]_{h}&& \mathbb{R}\\
\\
T^*Q/G\ar[uurr]_{h_G}
}
\]
where  $h_{G}\circ \pi=h$ is the natural projection of $h$. Consider the mapping $\Psi$ defined above
\[
\xymatrix{
T^*Q     \ar[dd]_{\pi}\ar[rr]^{h} && \mathbb{R} \\
\\
T^*Q/G    \ar[rruu]^{h_G} \ar@/_/[rr]_{\Psi}            &&
T^*(Q/G)\times_{Q/G}\tilde{\mathfrak{g}}^* \ar@/_/[ll]_{\Psi^{-1}} \ar[uu]_{h_G\circ\Psi^{-1}}
}
\]
and define $\tilde{h}_{\mu}:T^*(Q/G)\longrightarrow \mathbb{R}$ by $\tilde{h}_{\mu}(\tilde{\alpha}_{\tilde{q}})=\tilde{h}(\tilde{\alpha}_{\tilde{q}},[q,\mu])$,
where $\tilde{\alpha}_{\tilde{q}}\in  T^*_{\tilde{q}}(Q/G)$,
$\tilde{q}=[q]\in Q/G$, $\mu\in \mathfrak{g}^*$. Assume that $L$ is $G$-invariant solution of the Hamilton-Jacobi equation
for $h$ and define
\[
\tilde{L}=\Psi(\pi(L))\subset T^*(Q/G)\times_{Q/G}\tilde{\mathfrak{g}}^*.
\]
As we have proved before $\tilde{L}\subset T^*(Q/G)$ is a lagrangian submanifold with respect to $\omega_{Q/G}+\beta_{\mu}$. Using the previous results we can prove that a $G$-invariant solution for the Hamilton-Jacobi problem for
$h$ projects onto a solution of the Hamilton-Jacobi for
$\tilde{h}_{\mu}$. In addition, if $L$ is horizontal then $\tilde{L}$
is horizontal.

\begin{proposition}[Reduction]\label{reduction}
Given $L$ a $G$-invariant solution of the Hamilton-Jacobi equation, then $\tilde{L}$ is a solution of the Hamilton-Jacobi equation for
$\tilde{h}_{\mu}$ ($\mu=J(L)$). Moreover, if $L$ is horizontal, then $\tilde{L}$ is
horizontal.
\end{proposition}

\proof
Recall that since $L$ is $G$-invariant then $J(L)=\mu$. As we have seen before, $\tilde{L}=\pi(L)$ is a lagrangian submanifold
of $J^{-1}(\mu)/G$.  Now, we take $\mu\in\mathfrak{g}^*$ and since  it is a regular value of
$J$, then $J^{-1}(\mu)$ is a  submanifold of $T^*Q$. Since in our case, $L$ is $G$-invariant lagrangian submanifold then
$J$ is constant along $L$, say $J(L)=\mu$. Recall that $\mu\in\mathfrak{g}^*$ is a fixed point for $Ad^*$ if and only if $G_{\mu}=G$, and in this case $J ^{-1}(\mu)$ is
coisotropic. Therefore, we have that $\mu$ is such that
$G_{\mu}=G$. This happens for instance if $G$ is abelian. $\tilde{L}=\pi(L)$ is a lagrangian submanifold of $J^{-1}(\mu)/G$, but this
is a symplectic leaf with symplectic structure $\omega_{Q/G}+\beta_{\mu}$, when we are using the natural identification via $\Psi$ and considering
a fixed connection $A$ in $Q\rightarrow Q/G$ to obtain the corresponding decomposition. In addition, if $\tilde{\alpha}_{\tilde{q}}\in\tilde{L}$, then $\tilde{h}_{\mu}(\tilde{\alpha}_{\tilde{q}})=h(\Psi^{-1}(\tilde{\alpha}_{\tilde{q}},\mu))$. Therefore, $\tilde{h}_{\mu}$ is constant along $\tilde{L}$. Assume now that $L$ is horizontal, so $L=$Im$(\gamma)$, for a $1$-form $\gamma$ on $Q$ such that $d\gamma=0$. Since $\gamma$ takes values into $J^{-1}({\mu})$ and is $G$-invariant,
then $\gamma$ induces a mapping
\[
\xymatrix{
Q\ar[r] \ar@/_1pc/[rr]_{\tilde\gamma}& J^{-1}(\mu)\subset T^*Q/G\ar[r]^{\Psi}& T^*(Q/G)
}
\]
which is $G$-invariant. So it induces a new mapping $\tilde{\gamma}_{\mu}: Q/G\rightarrow T^*(Q/G)$
such that $\textrm{Im}(\tilde{\gamma}_{\mu})=\tilde{L}$.

\qed

We also prove a reconstruction theorem. With this theorem at hand, once a reduced solution is found it can be lifted to find a solution of the original unreduced problem.

\begin{proposition}[Reconstruction]\label{reconstruction} Assume that
  $\tilde{L}$ is a lagrangian submanifold of $(T^*(Q/G), 
  \omega_{Q/G}+\beta_{\mu})$ for some $\mu\in\mathfrak{g}^*$ wich is a
  fixed point of the Coadjoint action. Assume that $\tilde{h}_{\mu}$
  is the reduced hamiltonian defined as above and
that $\tilde{L}$ is a Hamilton-Jacobi solution for $\tilde{h}_{\mu}$. Using the diffeomorphism
\[
\xymatrix{
T^*Q/G \ar@/_/[rr]_{\Psi}&& T^*(Q/G)\times_{Q/G}\tilde{\mathfrak{g}}^* \ar@/_/[ll]_{\Psi^{-1}}
}
\]
we define $\hat{L}$ by 
\[
\hat{L}=\{(\tilde{\alpha}_{\tilde{q}},[\mu]_{\tilde{q}})\in
T^*(Q/G)\times_{Q/G}\tilde{\mathfrak{g}}^* \textrm{ such that }\tilde{\alpha}_{\tilde{q}}\in\tilde{L}\}
\]
and take
\[
L=\pi^{-1}(\hat{L}).
\]
Then
\begin{enumerate}
\item $L$ is $G$-invariant and lagrangian with respect to the canonical symplectic structure of the cotangent bundle, $\omega_Q$, and a solution for the Hamilton-Jacobi problem given by $h$.
\item If $\tilde{h}$ is horizontal, then $L$ is horizontal too.
\end{enumerate}
\end{proposition}

\proof
Since $\tilde{L}$ is a Lagrangian submanifold of $(T^*(Q/G),\omega_{Q/G}+\beta_{\mu})$ and $\Psi_{|J^{-1}(\mu)/G}$ is a symplectomorphism, then $\overline{L}=\Psi^{-1}(\hat{L})$ is a lagrangian submanifold of the symplectic leaf $J^{-1}(\mu)/G$. Since 
$\pi:T^*Q\rightarrow T^*Q/G$ is a submersion then $\pi^{-1}(\overline{L})$ is an immersed submanifold of dimension $\dim(\pi^{-1}(\overline{L}))=\dim(\overline{L})+\dim(G)$, and since 
\[
\begin{array}{ll}
\dim(\overline{L})&=\dim(\tilde{L})=1/2\cdot \dim(T^*(Q/G))
\\ \noalign{\medskip} 
&=1/2\cdot 2\cdot( \dim(Q)-\dim(G))=\dim(Q)-\dim(G),
\end{array}
\]
then $\dim(\pi^{-1}(\overline{L}))=( \dim(Q)-\dim(G))+\dim(G)=\dim(Q)$
which is half the dimension of $T^*Q$; so we only have to show that $\pi^{-1}(\overline{L})$ is an isotropic submanifold. Notice that since $J^{-1}(\mu)/G_{\mu}=J^{-1}(\mu)/G$, then the symplectic structure on $J^{-1}(\mu)/G$ (denoted by $\omega_{\mu}$) is the one obtained by the Marsden-Weinstein reduction theorem, which is characterized by the equation $i_\mu^*\omega_Q=\pi^*\omega_{\mu}$ where $i_{\mu}:J^{-1}(\mu)\rightarrow T^*Q$ is the  inclusion and $\omega_Q$ the canonical symplectic structure on $T^*Q$. Since $\pi^{-1}(\overline{L})\subset J^{-1}(\mu)$, it is easy to see that 
\[
(\omega_Q)_{|\pi^{-1}(\overline{L})}=(\pi^*{\omega_{\mu}})_{|\pi^{-1}(\overline{L})}=0,
\]
and we can conclude that $\pi^{-1}(\overline{L})$ is a lagrangian
submanifold. The fact that $h_{|\pi^{-1}(\overline{L})}=E$, where $E$
is a constant, follows from the identity
\[
h{|\pi^{-1}(\overline{L})}=(\tilde{h}_{\mu})_{|\tilde{L}}
\]
and thus the result holds.
\qed

\begin{remark}{\rm
It is clear that, by Propositions \ref{reduction} and
\ref{reconstruction}, we have a bijection  between $G$-invariant
solutions of Hamilton-Jacobi problem for $h$ and solutions of the
Hamilton-Jacobi equation for $\tilde{h}_{\mu}$ where $\mu$ is a fixed
point of the Coadjoint action.}
\end{remark}
\[
\xymatrix{
\left\{
\textrm{G-invariant solutions of HJ}
\right\}\ar[rr]^{\textrm{{\it one to one}}}
&&
\left\{
\textrm{reduced solutions of HJ}
\right\}\ar[ll]
}
\]
\begin{remark}{\rm
In the symplectic manifold $(T^*Q,\ \omega_Q)$, given a $1$-form  $\gamma$ on $Q$
its image is an horizontal lagrangian submanifold 
if and only if $d\gamma=0$. In that case that lagrangian submanifold is locally given by a generating
function $L=$Im($dS$). Given the symplectic manifold $(T^*(Q/G),\omega_{Q/G}+\beta_{\mu})$ it
is natural to ask which is the analogous condition to $d\gamma=0$. In \cite{hared}
one can check that $B_{\mu}$ is actually the pullback of a $2$-form on
the base $Q/G$ so $B_{\mu}=\pi_{Q/G}^*\beta_{\mu}$. So given a
$1$-form on $Q/G$, say $\gamma$, its image is lagrangian for the
modified structure if and only if
$0=\gamma^*(\omega_{Q/G}+\pi_{Q/G}^*\beta_{\mu})=d\gamma+\beta_{\mu}$
or equivalently $d\gamma=-\beta_{\mu}$. In that case, it is no possible
in general to find a generating function, and instead one PDE, we have a
system of algebraic-PDE equations.}
\end{remark}

%%%%%%%%%%%%%%%%%%%%%%%%%%%%%%%%%%%%%%%%%%%%%%%%%%%%%%%%%%%%%%%%%%%%%%%%%%%%%%%REDUCTION AND RECONSTRUCTION

\section{Reduction of H-J equation and reduction of
  dynamics}

Assume that we have a hamiltonian system $(T^*Q, \ \omega_Q, \ h)$ and let
$\gamma$ be a $1$-form which is a solution of the Hamilton-Jacobi equation
for $h$. Then we can construct the projected vector field $X_h^\gamma$
by
\[
X_h^\gamma=T\pi_Q\circ X_h\circ \gamma.
\]

A basic result in the Hamilton-Jacobi theory (see \cite{AM}) is that
$X_h^\gamma$ and $X_h$ are $\gamma$-related. If we assume that we are in the
conditions of the previous sections, that is, we have a free and
proper action $\Phi:G\times Q\rightarrow Q$ and all the constructions
previously introduced follow, we get a new (reduced) hamiltonian
system $(T^*(Q/G),\ \omega_{Q/G}+B_{\mu}, \ \tilde{h}_{\mu})$ and a
solution $\tilde{\gamma}$ of the corresponding Hamilton-Jacobi
theory.  As before, we can define the projected
vector field for the reduced system
\[
X_{\tilde{h}_{\mu}}^{\tilde{\gamma}}=T\pi_Q\circ X_{\tilde{h}_{\mu}}\circ \tilde{\gamma}.
\]
and therefore the current situation is the one described in the diagram below
\[
\xymatrix{
T^*Q \ar[d]&J^{-1}(\mu)\ar@{_
{(}->}[l]\ar[d]\\
Q\ar[rd]& J^{-1}(\mu)/G\ar[r]\ar[d]& T^*(Q/G)\ar[l]\\
& Q/G
}
\]
We point out the vector fields and the manifolsd on which they are
defined:
\[
\xymatrix{
T^*Q , \ X_h  \ar[d]^{\pi_Q} && T^*(Q/G), \ X_{\tilde{h}_\mu} \ar[d]^{\pi_{Q/G}}\\
Q, \ X_{h}^{\gamma} \ar@/^/[u]^{\gamma}&& Q/G, \ X_{\tilde{h}_\mu}^{\tilde{\gamma}}\ar@/^/[u]^{\tilde{\gamma}}
}
\]
The relation between the dynamics on $T^*Q$ and $J^{-1}(\mu)/G$
(recall that
we are identifying this space with $T^*(Q/G)$) is well-known. There
are reconstruction procedures to integrate the vector field $X_h$
after integrating the vector field $X_{\tilde{h}_\mu}$. So we have
\begin{figure}[h]
\[
\xymatrix{
T^*Q , \ X_h  \ar[d] \ar@/^/[rr]^{\textrm{projection}}&& T^*(Q/G), \ X_{\tilde{h}_\mu} \ar[d]\ar@/^/[ll]^{\textrm{reconstruction}}\\
Q, \ X_{h}^{\gamma} \ar[u]^{\textrm{Hamilton-Jacobi}}&& \ar@{.>}[ll]^{\textrm{reconstruction}} Q/G, \ X_{\tilde{h}_\mu}^{\tilde{\gamma}}\ar[u]_{\textrm{Hamilton-Jacobi}}
} \]\caption{Relations between vector fields}\label{diagram5}
\end{figure}

\noindent Moreover, since $\tilde{\gamma}\circ\pi_G=\pi\circ\gamma$ we can
conclude that the vector field $X_{h}^{\gamma}$ projects onto $
X_{\tilde{h}_\mu}^{\tilde{\gamma}}$ via $\pi_G$, and so, $X_{h}^{\gamma}$ is
$G$-invariant.

We recall now the basic reconstruction procedure to
integrate the vector field $X_{h}^{\gamma}$ via the integration of $
X_{\tilde{h}_\mu}^{\tilde{\gamma}}$  in order to complete the diagram in Figure
\ref{diagram5}. Let $c:(a,b)\subset \mathbb{R}\rightarrow Q/G$ be an
integral curve of $X_{\tilde{h}_\mu}^{\tilde{\gamma}}$  and consider a
curve $d(t):(a,b)\rightarrow Q$ such that $\pi_{Q/G}\circ d=c$; for
instance, since we made the previous constructions using a connection
on the principal bunle $\pi_G:Q\rightarrow Q/G$ then $d$ can be taken
as the horizontal lift of $c$. Next, consider the connection $1$-form,
that we will denote also by $A:TQ\rightarrow \mathfrak{g}$, and assume
that we have a curve $g:(a,b)\rightarrow G$ such that
$\frac{d}{dt}g(t)=A(X_{h}^{\gamma}(d(t))-\frac{d}{dt}d(t))$ where we
are using the identification $TG\equiv G\times \mathfrak{g}$ given by
the left trivialization. It is easy to check that then $g(t)\cdot{}
d(t)$ is an integral curve of $X_{h}^{\gamma}$.

%%%%%%%%%%%%%%%%%%%%%%%%%%%%%%%%%%%%%%%%%%%%%%%%%%%%%%%%%%%%%%%%%%%%%%%%%%%%%%%%%%%
%%                           EXAMPLES
%%%%%%%%%%%%%%%%%%%%%%%%%%%%%%%%%%%%%%%%%%%%%%%%%%%%%%%%%%%%%%%%%%%%%%%%%%%%%%%%%%%

\section{Examples}\label{examples}
It is our believe that the theory above described has wide applicability in concrete situations. Here we present some examples but we would like to stress that much more involved settings fall in our setting.

\subsection{Lie groups} Let $G$ be a Lie group and $T^*G$ its cotangent bundle. Using left trivialization we have the identification  $T^*G\cong G\times\mathfrak{g}^*$. Since $T^*G/G\cong G\times\mathfrak{g}^*/G\cong\mathfrak{g}^*$ then, to find a $G$-invariant solution $L$ of the Hamilton-Jacobi problem is equivalent to finding an element $\mu\in\mathfrak{g}^*$ such that $Ad^*_g(\mu)=\mu$ for all $g\in G$. Given such $\mu$ we can construct $L\subset G\times \mathfrak{g}^*$ given by $L=G\times \{\mu\}$. It is easy to see that a $1$-form defined in this way is closed, $G$-invariant and satisfies $H_{|G\times \{\mu\}}=\tilde{H}(\mu)$. Therefore we obtain a characterization of the closed $G$-invariant $1$-forms on a Lie group.

\subsection{ The trivial case: $Q=M\times G$}

Assume now that we have $Q=M\times G$ and we are considering the
action
\[
\begin{array}{rccl}
\Phi: &G\times (M\times G)& \longrightarrow & (M\times G)\\ \noalign{\medskip}
& (g,(m,h))&\rightarrow& (m,g\cdot h).
\end{array}
\]
If we trivialize $T^*G=G\times \mathfrak{g}^*$ via the left action,
then the lifted action, $\Phi^{T^*}$ is given by

\[
\begin{array}{rccl}
\Phi^{T^*}: &G\times (T^*M\times G\times \mathfrak{g}^*)& \longrightarrow & (T^*M\times G\times \mathfrak{g}^*)\\ \noalign{\medskip}
& (g,(\alpha_m,h,\mu))&\rightarrow& (\alpha_m,g\cdot h,\mu).
\end{array}
\]
The momentum map is given by $J(\alpha_m,g,\mu)=Ad^*_{g^{-1}}\mu$. If we have the hamiltonian system $(T^*(M\times G), \ H , \
\Omega_{M\times G})$ (with $H$ assumed $G$-invariant), given $\mu$ such that
$G_{\mu}=G$ then $J^{-1}(\mu)/G\cong T^*M$
and $\tilde{H}_{\mu}(\alpha_m)=H(\alpha,g,\mu)$
where by the $G$-invariance of $H$ the element $g$ is arbitrary. In this case, the reduced system is equivalent to the hamiltonian
system given by $(T^*M, \ \tilde{H}_{\mu}, \ \Omega_M)$. Assume that
$S_M:T^*M\rightarrow \mathbb{R}$ is the generating function of
$\tilde{L}$, a horizontal lagrangian submanifold which solves the
Hamilton-Jacobi problem. On the other hand it is easy  to
see that $\mu$ viewed as a section of the projection onto $G$,
$G\times \mathfrak{g}^*\rightarrow G$, is a closed $1$-form and so
there exists $S_{G}:G\rightarrow \mathbb{R}$ such that
Im$(dS_{G})=(g,\mu)$. Let us denote by $S_{M\times G}$ the generating
function of the corresponding lagrangian submanifold $L$ obtained by
reconstruction from $\tilde{L}$,  then we have.

\begin{lemma}
The  generating
functions are related by
\[
S_{M\times G}=S_M+S_G+c
\]
where $c$ is a constant on each connected component.
\end{lemma}
\proof Given $\xi\in \mathfrak{g}$, since Im$(dS_{M\times G})\subset
J^{-1}(\mu)$ then $dS(\xi_{M\times Q})=\mu(\xi)=d(S_M+S_G)(\xi_{M\times Q})=dS_G(\xi_{M\times Q})$. Given
$X\in T_mM$ the analogous computations holds and the result follows.

\qed

\subsubsection{Time-dependent H-J solution for time-independent systems}\label{time-dependentfortimeindependent}

An immediate application of the previous result is the obtainment of the
classical relation between time-dependent and time independent
solutions of the Hamilton-Jacobi equation. This is a very classical ansatz that follows from our results.

Let be $H:T^*Q\rightarrow \mathbb{R} $ and consider the corresponding
hamiltonian $H^{\mathbb{R}}=H\circ p_{T^*Q}+e:T^*(\mathbb{R}\times
Q)\rightarrow \mathbb{R}$, where $ p_{T^*Q}:T^*(\mathbb{R}\times
Q)\rightarrow T^*Q$ is the projection onto $T^*Q$ and $e$ denotes the time conjugate momentum. We can introduce the action given by translation in time
\[
\begin{array}{rccl}
\Phi: &\mathbb{R}\times (\mathbb{R}\times Q)& \longrightarrow & (\mathbb{R}\times Q)\\ \noalign{\medskip}
& (r,(t,q))&\rightarrow& (t+r,q).
\end{array}
\]
The corresponding lifted action is
\[
\begin{array}{rccl}
\Phi^{T^*}: &\mathbb{R}\times T^*(\mathbb{R}\times Q)& \longrightarrow
&T^* (\mathbb{R}\times Q)\\ \noalign{\medskip}
& (r,(t_,e,\alpha_q))&\rightarrow& (t+r,e,\alpha_q).
\end{array}
\]
 The momentum map is just
\[
J(t,e,\alpha_q)=e.
\]
If $E\in\mathbb{R}\cong\mathbb{R}^*$ then $\mathbb{R}_{E}=\mathbb{R}$
since the group is abelian and $J^{-1}(E)\cong
\mathbb{R}\times T^*Q$ and $\mathbb{R}\times T^*Q/\mathbb{R}\cong
T^*Q$. Summarizing, we have that $(J^{-1}(E)/G, \ \tilde{H^{\mathbb{R}}}_{E},
  \tilde{\Omega})$ is given by $(T^*Q, \ H, \ \Omega_Q)$ and, if we
  denote by $S$ the generating function of $L$ and by $W$ the
  generating function of $\tilde{L}$, then we obtain $S_{\mathbb{R}}=t\cdot E$ and $S_{Q}=W$ and we recover
\[
S=t\cdot E+W
\]

%%%%%%%%%%%%%%%%%%%%%%%%%%%%%%%%%%%%%%%%%%%%%%%%%%%%%%%%%%%%%%%%%%%%%%%%%%%%%%%%%%%%%%%
\subsection{Complete Solutions}\label{completesolution}

This subsection is devoted to applying the previous results to what is usually called a complete solution of the
Hamilton-Jacobi equation. The knowledge of  a complete solution of the
Hamilton-Jacobi equation is equivalent to integrating the Hamilton's
equations of motion (see \cite{Arnold}). Before getting into our results, we sketch in
this subsection the classical results. They are local and written in a
coordinate dependent way, but the global, geometric
aspects of the theory are easier to understand after taking a look at
the classical theory. We restrict ourselves to the time-independent case but the results can be easily extended to the time-dependent setting.

Let $h(q^i,p_i)$ be a hamiltonian on the phase space
$(q^i,p_i)$, $i=1,\ldots,n$. By a complete solution of the Hamilton-Jacobi equation
for $h$ we mean the following.

\begin{definition} A complete solution of the Hamilton-Jacobi equation
  for the hamiltonian $h(q^i,p_i)$, $i=1,\ldots,n$ is a real-valued
  function $S(t,q^i,\alpha^i)$, $i=1,\ldots,n$, such that
\begin{enumerate}
\item For every  (fixed) value of the parameters $(\alpha^i)$,
  $S(t,q^i,\alpha^i)$ satisfies the Hamilton-Jacobi equation,
\[
\displaystyle\frac{\partial S}{\partial t}+h(t,q^i,\displaystyle\frac{\partial S}{\partial q^i}(t,q^i))=0.
\]
\item The non-degeneracy condition: consider the matrix with component
  $i,j$ given by $\displaystyle\frac{\partial^2 S}{\partial q^i\partial
    \alpha^j}$, that we denote by $(\displaystyle\frac{\partial^2 S}{\partial q^i\partial
    \alpha^j})$, then \[det(\displaystyle\frac{\partial^2 S}{\partial q^i\partial
    \alpha^j})\neq 0.\]
\end{enumerate}
\end{definition}
Then, we can define (at least locally by the implicit function theorem) the following implicit, time-dependent transformation, from the $(t,q^i,p_i)$-space to the $(t,\alpha^i,\beta_ i)$-space:
\begin{equation}\label{transformation}
\begin{array}{lr}
\displaystyle\frac{\partial S}{\partial q^i}(t,q^i,\alpha^i)=p_i & -\displaystyle\frac{\partial S}{\partial \alpha^i}(t,q^i,\alpha^i)=\beta_i.
\end{array}
\end{equation}
A computation shows that this transformation sends the system to {\it equilibrium}, i.e., Hamiltlon's equations become now
\begin{equation}\label{equilibrium}
\begin{array}{c}
\displaystyle\frac{d \alpha^i}{dt}(t)=0,
\\ \noalign{\medskip}
\displaystyle\frac{d \beta_i}{dt}(t)=0,
\end{array}
\end{equation}
see \cite{AM, Arnold}.

We give now a geometric interpretation of the previous
procedure. The function $S$ can be interpreted as a function on the
product manifold $\mathbb{R}\times Q\times Q$ and so $\im(dS)$ is a
lagrangian submanifold in $\trqq$ (notice that we are thinking about the
$(q^i)$ as coordinates on the first $Q$, and $(\alpha^i)$ as coordinates on the
second factor $Q$). On the other hand, consider the
projections $\pi_I:\trqq\rightarrow \mathbb{R}\times T^*Q$, $I=1,2$,
defined by $\pi_I(t,e,\alpha^1,\alpha^2)=(t,(-1)^{I+1}\alpha^I)$. With these
geometric tools, the non-degeneracy condition is equivalent to
saying that ${\pi_I}_{|\im(dS)}$ is a local diffeomorphism for
$I=1,2$. We assume here for simplicity that it is a global
diffeomorphism, so we can consider the mapping $
 {\pi_2}_{|\im(dS)}\circ ({\pi_1}_{|\im(dS)}) ^{-1}:\trq\rightarrow
 \trq$. This mapping can be easily checked to be the global
 description of the change of variables introduced
 above. The Hamilton-Jacobi equation, can be
 understood as the fact that $dS^*h^{ext}=0$, where
 $h^{ext}=\pi_1^*h+e$. The diagram below helps to have a global
 picture of the procedure:
\begin{figure}[h]
\[
\xymatrix{
&&\im(dS)\subset T^*(\mathbb{R}\times Q\times
Q)\ar[r]^>>>{\pi_1^*h+e\qquad }\ar[dl]^{\pi_1}
\ar[dr]_{\pi_2}\ar[dd]_>>>{\pi_{\mathbb{R}\times Q\times Q}}&\mathbb{R}\\
\mathbb{R}&\mathbb{R}\times T^*Q\ar[l]^{H}\ar[rr]_<<{\qquad \qquad \qquad\qquad {\pi_2}\circ ({\pi_1}_{|\im(dS)}) ^{-1}}|\hole\ar[dd]_{id_{\mathbb{R}}\times \pi_Q}
\ar@/^/[ru]^<<{({\pi_1}_{|\im(dS)}) ^{-1} }&&\mathbb{R}\times T^*Q
\ar[dd]^{id_{\mathbb{R}}\times \pi_Q}\\
&&\mathbb{R}\times Q\times Q \ar@/_/[uu]_>>>>>>>{dS}|\hole\ar[dl]_{pr_1}\ar[dr]^{pr_2}& \\
&\mathbb{R}\times Q&&\mathbb{R}\times Q
}\]\caption{Geometric interpretation of complete solutions of the H-J equation}\label{diagram}
\end{figure}

In the precedent setting all the information is given by the lagrangian manifold defined
by $\im(dS)$, so we can introduce a generalized solution to the
Hamilton-Jacobi equation as follows.
 
\begin{definition}
A lagrangian submanifold $L$ in $\trqq$ is a complete solution of the
Hamilton-Jacobi equation if
\begin{enumerate}
\item $L\subset ({h^{ext}})^{-1}(e)$.
\item The restriction of $\pi_I$ to $L$ is a diffeomorphism. From now on we will refer to this property as the non-degeneracy condition.
\end{enumerate}
\end{definition}

\begin{remark}{\rm
When $L$ is given by $\im(dS)$ we say that $S$ is a generating
function for the transformation induced by $L$. This type of
generating functions are usually called in the literature {\it type I
generating functions}, see \cite{Goldstein}. It is remarkable that our
theory deals with the lagrangian submanifolds instead of their
generating functions, so our theory is applicable to other types of
generating functions. This does not happen in previous approaches to
reduction of 
the Hamilton-Jacobi theory.}
\end{remark}

Under the previous conditions we are still able to define the
symplectomorphism that solves Hamilton's equations. We can now apply our reduction procedure. Assume that we have an
action $\Phi:G\times Q\rightarrow Q$, such that $\Phi^{T^*}$ leaves the hamiltonian
invariant. We consider the diagonal
action
\[
\begin{array}{rccl}
\Phi_0:&G\times \mathbb{R}\times Q\times Q &\rightarrow &\mathbb{R}\times
Q\times Q \ \\ \noalign{\medskip}
& (g,(t,q^1,q^2))&\rightarrow & (t,\Phi(g,q^1),\Phi(g,q^2)).
\end{array}
\]
It is easy to see that $\Phi^{T^*}_0$ leaves $h^{ext}$
invariant and the corresponding momentum mapping is
$J_0=J\circ\pi_1-J\circ \pi_2$, where $J$ is the momentum mapping
corresponding to the action $\Phi$. Then, we can look for $G$-invariant complete
solutions. After applying our reduction method, we obtain the (reduced)
cotangent manifold  $T^*(\mathbb{R}\times
\frac{Q\times Q}{G})$. All the reduction theory for Hamilton-Jacobi applies in this manner to complete solutions in a straightforward way.

\begin{remark}{\rm
A simple computation, following  the arrows in Figure \ref{diagram}, shows that $G$-invariant lagrangian submanifolds in
$\trqq$ which satisfy the non-degeneracy condition induce
time-dependent $G$-equivariant symplectic automorphisms on $T^*Q$.}
\end{remark}

\begin{remark}{\rm
 If our hamiltonian comes from a regular lagrangian, $\mathcal{L}$, then there
 is always a (local) $G$-invariant solution which lives in
 $J_0^{-1}(0)$, just the one given by the action functional
\[
S(t,q,\overline{q})=\int_c \mathcal{L}(\dot{c})\ dt,
\]
where $c: [a,b]\rightarrow Q$ is the curve satisfying the Euler-Lagrange equations and verifying $c(a)=q$ and $c(b)=\overline{q}$. Under the previous assumptions, that curve exists for $q$ and $\overline{q}$ close enough.}
\end{remark}

\begin{remark}{\rm There is a very important lagrangian submanifold, the one given by the flow. Let $\Psi_t^h$ be the flow of the hamiltonian vector field $X_h$. Then we have the lagrangian submanifold in $T^*(\mathbb{R}\times Q\times Q)$
\[
L=\{(t,h(t,\alpha_q),\alpha_q,-\Psi^h_t(\alpha_q))\textrm{ such that }t\in \mathbb{R}, \ \alpha_q\in T^*Q\}
\]
At the end, the Hamilton-Jacobi theory is about finding a generating function for this lagrangian submanifold. If $G$ is a symmetry of the hamiltonian then $L$ is $G$-invariant and lives in the $0$-level set, as a consequence of the conservation of the momentum mapping. This lagrangian submanifold can locally be obtained by type II generating function (see next section).}

\end{remark}

\begin{remark}{\rm
Observe that $T^*(\mathbb{R}\times
\frac{Q\times Q}{G})$ has a well-known geometric structure; indeed, it is the cotangent bundle of the gauge groupoid $\mathbb{R}\times
\frac{Q\times Q}{G}$. It suggest that the geometric structure behind all this theory is the symplectic groupoid structure. Moreover, following this pattern we were able to develop a Hamilton-Jacobi theory for certain Poisson manifolds that will appear in a forthcomming paper \cite{hamiltonjacobicompleto}.}
\end{remark}

\begin{remark}{\rm
The reduced lagrangian submanfold $\hat{L}\subset T^*(\mathbb{R}\times
\frac{Q\times Q}{G})$ induces a (Poisson) transformation
$\mathbb{R}\times T^*Q/G\rightarrow \mathbb{R}\times T^*Q/G$, using
the source and the target of the groupoid structure, in the same way
we have used the projections $\pi_I$ above. The Poisson structure
considered on $\mathbb{R}\times T^*Q/G$ is the product of the $0$
Poisson structure on $\mathbb{R}$ and the natural Poisson structure
induced on $ T^*Q/G$ by the quotient of the symplectic structure on $T^*Q$. In the case $Q=G$, the source and the target are the left and right momentum mappings  $J^L$ and $J^R$, and in the pair groupoid case the projections $\pi_I$. This reinforces the idea that symplectic groupoids play an essential role in this theory.}
\end{remark}

\begin{remark}{\rm As a by-product, we obtain all the results related to
  the reduction of the Hamilton-Jacobi theory of reference
  \cite{Ge}. The previous discussion specializes to Lie groups, $Q=G$
  and then the reduced space $T^*(\mathbb{R}\times
\frac{Q\times Q}{G})=T^*(\mathbb{R}\times
\frac{G\times G}{G})$ can be identified with $T^*(\mathbb{R}\times
G)$ to recover the theory in Ge and Marsden, \cite{Ge}.}
\end{remark}

%%%%%%%%%%%%%%%%%%%%%%%%%%%%%%%%%%%%%%%%%%%%%%%%%%%%%%%%%%%%%%%%%%%%%%%%%%%%%%%%%%%%%%%
\subsubsection{Other types of generating function}

The goal of this section is to show how our results can be applied to other types of generating functions; in this way we recover some classical results about cyclic coordinates. We chose the so-called {\it type II} generating functions, but since our theory is valid for any lagrangian submanifold it can be used to deal with any type of generating functions. This type II generating functions are very important, because they can generate the identity transformation and all the ``nearby''  canonical transformations. We introduce below the classical situation,  we assume that $Q=\mathbb{R}^n$
and so $T^*Q=\mathbb{R}^{2n}$ and consider global coordinates
$(q^i,p_i)$, $i=1,\ldots,n$. Doubling these coordinates we get a coordinate system for
$T^*(Q\times Q)=\mathbb{R}^{4n}$, say $(q^i,p_i,\alpha^i,\beta_i)$, and we obtain coordinates $(t,e,q^i,p_i,\alpha^i,\beta_i)$ on $T^*(\mathbb{R}\times Q\times Q)$. Given a function $S(t,q^i,\beta_i)$ it is easy to check that the submanifold given by
\[
L=\{(t,\displaystyle\frac{\partial S}{\partial t}(t,q^i,\beta_i),q^i,\frac{\partial S}{\partial q^i}(t,q^i,\beta_i),\frac{\partial S}{\partial \beta_i}(t,q^i,\beta_i),-\beta_i)\textrm{ such that }t, \ q^i,\ \beta_i\in\mathbb{R}\}
\]
is lagrangian.

\begin{remark}{\rm A more detailed explanation about the construction of this submanifold can be found in \cite{hamiltonjacobicompleto}.} 
\end{remark}

Following the same pattern than above, such generating function gives a time-dependent canonical transformation, given implicitly by
\begin{equation}\label{coordinatechangenonfree}
\begin{array}{cc}
\displaystyle\frac{\partial S}{\partial q^i}(t,q^i,\beta_i)=p_i, &\displaystyle\frac{\partial S}{\partial \beta_i}(t,q^i,\beta_i)=\alpha^i.
\end{array}
\end{equation}
as long as det$(\displaystyle\frac{\partial^2 S}{\partial q^i\partial
    \beta_j})\neq 0$.
		
		Now, our reduction procedure can be applied to the lagrangian submanifold $L$ in a straightforward way. We work out here the details in the case of a time-independent hamiltonian with one cyclic variable in order to recover some results present in the literature, the cases with more than one cyclic variables are obvious. Assume that $h(q^i,p_i)$ does not depend on $t$ and $q^1$, {\it i.e.} $q^1$ is a cyclic variable. We are looking for a type II solution of the Hamilton-Jacobi equation for $h$, that is, $S(t,q^i,\beta_i)$ such that
		\begin{enumerate}
		\item The Hamilton-Jacobi equation $\displaystyle\frac{\partial S}{\partial t}+h(q^i,\frac{\partial S}{\partial q^i})=E$, where $E$ is a real constant.
		\item Non-degeneracy condition, det$(\displaystyle\frac{\partial^2 S}{\partial q^i\partial
    \beta_j})\neq 0$.
		\end{enumerate}
		Using Section \ref{time-dependentfortimeindependent} we assume $S(t,q^i,\beta_i)=t\cdot E+W(q^i,\beta_i)$, where $W$ should satisfy
		\begin{equation}\label{hamiltonjacobilast}
		h(q^i,\frac{\partial W}{\partial q^i}(q^i,\beta_i))=F
		\end{equation}
		for some  constant $F$ and the non-degeneracy condition. Notice that such function $W$ gives a lagrangian submanifold in $\mathbb{R}^{4n}$ by
		\[
L_1=\{(q^i,\frac{\partial W}{\partial q^i}(q^i,\beta_i),\frac{\partial W}{\partial \beta_i}(q^i,\beta_i),-\beta_i)\textrm{ such that } q^i,\ \beta_i\in\mathbb{R}\}.
\]
In order to solve \eqref{hamiltonjacobilast} we use the theory previouly developed. Notice that $q^1$ is a cyclic variable if and only if the hamiltonian is invariant by the $\mathbb{R}$ action given by $(r,(q^1,\ldots,q^i,\ldots,q^n))=(q^1+r,\ldots,q^i,\ldots,q^n)$, which has an associated momentum mapping given by $J(q^i,p_i)=p_1$. The corresponding diagonal action is given by
		\[
		(r,(q^i,p_i,\alpha^i,\beta_i))=(q^1+r,\ldots,q^i,\ldots,q^n,p_i,\alpha^1+r,\ldots,\alpha^i,\ldots,\alpha^n,\beta_i))
		\]
		with momentum mapping 
		\[
		J(q^i,p_i,\alpha^i,\beta_i)=p_1+\beta_1.
		\]
		So, if we are looking for a lagrangian submanifold $L_1$ living in the $0$ level set of the momentum mapping, that is natural regarding the previous remarks, we should impose 
		\[
		\displaystyle\frac{\partial W}{\partial q^1}-\beta_1=0
		\]
		which implies,  by simple integration, that
		\[
		W=q^1\beta_1+V(q^i,\beta_j), \quad i=2,\ldots,n; \ j=1,\ldots,n,
		\]
		where the important observation here is that $V$ does not depend on the cyclic variable $q^1$. In this way, we have reduced the number of independent variables by one, this could simplify drastically the Hamilton-Jacobi equation. Here we have recovered the classical ansantz for cyclic variables, see \cite{Ardema,Goldstein}, from our geometric interpretation of the Hamilton-Jacobi theory in a straightforward way.
		\begin{remark} In the case of more than one cyclic variables an analogous result holds, there
			\[
		W=q^l\beta_l+V(q^i,\beta_j), \quad i=k,\ldots,n; \ j=1,\ldots,n,
		\]
		where $l=1,\ldots k$ are the cyclic variables.
		\end{remark}
		
		We show how to obtain a complete solution, using this method, of the Hamilton-Jacobi equation for a heavy-top with to equal moments of inertia. The hamiltonian is given by
		\[
		h(\theta, \phi,\psi,p_\theta,
                p_\phi,p_\psi,)=1/2\left(\displaystyle\frac{p_\theta^2}{I}+\frac{(p_\phi-p_\psi
                    \cos(\theta))^2}{I
                    \sin^2(\theta)}+\frac{p_\psi^2}{J}\right)+mgl \cos(\theta),
		\]
		where $I,\ J$ are the moments of inertia, $m$ the mass, $g$ the acceleration of gravity. Using the constructions above 
		\[
		S=t\cdot E+W(\theta, \phi,\psi,\beta_1,\beta_2,\beta_3)
		\]
		and 
		\[
		W(\theta, \phi,\psi,\beta_1,\beta_2,\beta_3)=\phi\cdot \beta_2+\psi\cdot \beta_3+V(\theta,\beta_1,\beta_2,\beta_3).
		\]
		Taking into account all this expressions, we get for the Hamilton-Jacobi equation
		\[
1/2\left(\displaystyle\frac{\partial
    V^2}{\partial\theta}\frac{1}{I}+\frac{(\beta_2-\beta_3
    \cos(\theta))^2}{I \sin^2(\theta)}+\frac{\beta_3^2}{J}\right)+mgl \cos(\theta)=F.		
		\]
		
		From here it is immediate to integrate the equation and to chose a solution that is non-degenerate; notice that the only unknown is $\frac{\partial V}{\partial \theta}$ and so by simple integration we can achieve the solution. Although this result was well-known classically, our point here is that it fits directly within our setting. Compare our results with \cite{Ardema}, p. $315$.
%%%%%%%%%%%%%%%%%%%%%%%%%%%%%%%%%%%%%

\subsection{Calogero-Moser system}

We would like to treat another concrete application. Here we deal with a Calogero-Moser system of two
particles. Although simple, this system illustrates how our method
works. Consider the hamiltonian
\[
\begin{array}{rccl}
H: & T^*\mathbb{R}^2&\rightarrow &\mathbb{R}\\ \noalign{\medskip}
&(q^1,q^2,p_1,p_2)&\rightarrow &H (q^1,q^2,p_1,p_2)=1/2\ (p_1^2+p_ 2^2)+1/(q^1-q^2)^2.
\end{array}
\]
In this example $Q=\mathbb{R}^2$ and the action
\[
\begin{array}{rccl}
\Phi: & \mathbb{R}\times \mathbb{R}^2&\rightarrow & \mathbb{R}^2\\ \noalign{\medskip}
&(r,(q^1,q^2,p_1,p_2))&\rightarrow& (r+q^1,r+q^2)
\end{array}
\]
is a symmetry of the system, i.e., the hamiltonian is invariant under
the corresponding lifted action, $\Phi^{T^*}$. We are now looking for a
solution of the equation
\[
H\left(q^1,q^2,\displaystyle\frac{\partial W}{\partial
  q^1},\displaystyle\frac{\partial W}{\partial q^2}\right)=E
\] 
which in this case becomes
\[
1/2\ \left(\displaystyle\frac{\partial W}{\partial q^1}^2+\displaystyle\frac{\partial W}{\partial
  q^2}^2\right) +1/(q^1-q^2)^2=0.
\]
We are looking for solutions in $J^{-1}(0)$, where 
\[
\begin{array}{rccl}
J: & T^* \mathbb{R}^2&\rightarrow & \mathbb{R}\\ \noalign{\medskip}
&(q^1,q^2,p_1,p_2)&\rightarrow &J(q^1,q^2,p_1,p_2)=p_1+p_2
\end{array}
\]
 Then, $J^{-1}(0)=\{(q^1,q^2,p_1,p_2)\textrm{ such that }p_2=-p_1\}$,
 and thus coordinates on $J^{-1}(0)$ are given by
 $(q^1,q^2,p)\rightarrow (q^1,q^2,p,-p)$. In the same way,
 $J^{-1}(0)/\mathbb{R}$ is $\mathbb{R}^2$, with coordinates $(q,p)$ and
 the natural projection $\pi:J^{-1}(0)\rightarrow J^{-1}(0)/\mathbb{R}$
 reads $\pi(q^1,q^2,p)=(q=q^1-q^2,p)$. Some abuse of notation is made,
 but there is no room for confusion. Since $H$ is
 $\mathbb{R}$-invariant there is a reduced hamiltonian,
 $\overline{H}:J^{-1}(0)/\mathbb{R}\equiv \mathbb{R}^2\rightarrow
   \mathbb{R}$, such that $\overline{H}(q,p)= p^2+1/q$. Now the reduced
  Hamilton-Jacobi equation is just an ODE
\[
\overline{H}\left(q,\displaystyle\frac{\partial \overline{W}}{\partial q}\right)=E,
\]
and the reduced Hamilton-Jacobi equation can be integrated looking
for a primitive
\[
\overline{W}=\int (E-1/q^2)^{1/2}
\]
for the values of $q$ where it makes sense.  That can be checked to be
\[
\overline{W}(q)=(\sqrt{E q^2-1}-\arctan(\displaystyle\frac{1}{\sqrt{E
    q^2-1}})). %\sqrt{2}
\]
Then, the reconstruction
procedure gives us
\[
W(q^1,q^2)=\overline{W}(q^1-q^2)=(\sqrt{E (q^1-q^2)^2-1}-\arctan(\displaystyle\frac{1}{\sqrt{E (q^1-q^2)^2-1}}) ) %\sqrt{2},
\]
which is defined when $q^1-q^2>\sqrt{E}$.

\section{Conclusions and Future Research}

In this paper we developed a complete theory of reduction and reconstruction of the Hamilton-Jacobi equation for hamiltonian systems with symmetry. The symmetry is supposed to be the lifted action of an action on the configuration manifold, $Q$. Our theory is explained for the time-independent and time dependent Hamilton-Jacobi equations, moreover, complete solutions are also considered. We showed that our theory unifies and extends previous approaches by Ge and Marsden and we can recover in a straightforward way the classical ansatz used in the literature to deal with cyclic variables and time-independent hamiltonians. The results in \cite{GeEquivariant} are also particular instances of our approach. On the other hand, one of the main points of our theory is that we link reduction theory with symplectic groupoids. That link was started in \cite{GeHJ} but our approach is quite different and will appear elsewhere (\cite{hamiltonjacobicompleto}) with some applications to (Poisson) numerical methods. Some open problem related to this work are:

\begin{enumerate}
\item {\it Relate our theory to the theory of generating functions in \cite{MeyerI,MeyerII}.} The theory developed there relies on generating function, so it seems that our theory should be the natural framework to deal with this kind of theories. Connections with the {\it Poincar\'e generating function} would be also very interesting.

\item {\it Develop an analogous theory for general symmetries.}
  Although quite useful out setting only deals at this moment with
  cotangent lifts of symmetries, to develop an analogous theory for
  any kind of symmetries should provide means to integrate more
  general systems. The results in \cite{Ortega} could be of some help
  in this regard.

\item {\it Construction of geometric integrators from complete
    solutions of the \linebreak Hamilton-Jacobi equation.} Complete solutions
  of the Hamilton-Jacobi equations are sometimes hard to find, but
  they can be approximated in order to find numerical methods that
  preserve the underlying geometry. This procedure is well-known in the
  symplectic case, see \cite{Kang,Hairer}. 
Our setting is useful in order to
  develop the analogous Poisson integrators in the situations treated here. Related work will appear in \cite{hamiltonjacobicompleto}.
\end{enumerate}

%%%%%%%%%%%%%%%%%%%%%%%%%%%%%%%%%%%%%%%%%%%%%%%%%APENDICES

\begin{appendices}

\section{Principal bundles and adjoint bundles}

Consider the $G$-principal bundle \[\pi:Q\rightarrow Q/G\] with the action on the left \[\Phi:G\times Q\rightarrow Q\]
and $F$ a manifold endowed  with a left action \[\rho:G\times F\rightarrow F.\] 
We shall construct the fiber bundle  $Q\times_G F$. Let $Q\times F$ be the product manifold and we introduce the action
\[
\begin{array}{rcl}
G\times (Q\times F):&\longrightarrow&Q\times F \\ \noalign{\medskip}
(g,(q,f))&\rightarrow&(\Phi(g,q),\rho(g,f)).
\end{array}
\]
The quotient space of $Q\times F$ by this action is called the the
fiber bundle over the base $Q/G$ with standard fiber $F$ and structure
group $G$, which is associated with the principal bundle $Q$ and it is
denoted by $Q\times_G F$. We introduce now the differentiable structure of this bundle. The mapping 
\[
\begin{array}{rccl}
\tilde{\pi}:& Q\times F&\longrightarrow &Q/G \\ \noalign{\medskip}
&(q,f)&\rightarrow &\pi(q)
\end{array}
\]
induces a mapping $\hat{\pi}:Q\times_G F\rightarrow Q/G$. Since for
each $x\in Q/G$ there exists a neighborhood $U$ such that
$\pi^{-1}(U)\equiv U\times G$, it can be easily seen that there is an
isomorphism $\hat{\pi}^{-1}(U)\equiv U\times F$. Therefore we can
introduce a differentiable structure on $Q\times_G F$ by the
requirement that $\hat{\pi}^{-1}(U)$ is an open submanifold of
$Q\times_G F$ diffeomorphic with $U\times F$ under the isomorphism
$\hat{\pi}^{-1}(U)\equiv U\times F$. Then, it follows that $\hat{\pi}$ is a
differentiable mapping.

We now specialize the previous construction to
the case when $F=\mathfrak{g^*}$ and the action is given by
\[
\begin{array}{rccl}
\tilde{\rho}: & G\times \mathfrak{g^*}&\longrightarrow & \mathfrak{g}^* \\ \noalign{\medskip}
&(g,\mu)& \rightarrow &\tilde{\rho}(g,\mu)=Ad^*_{g^{-1}}(\mu)
\end{array}
\]
The corresponding bundle obtained using the action
$\tilde{\rho}$  will be denoted by $\tilde{\mathfrak{g}}^*$.

\section{Magnetic Terms}\label{B}

Let $Q$ be a manifold, $\Phi:G\times Q\rightarrow Q$ a free and proper
action, $\Phi^{T^*}$ the cotangent-lifted action and
$J:T^*Q\rightarrow \mathfrak{g}$ the
corresponding momentum mapping. Recall that $Q/G$ is endowed with a 
structure of differentiable manifold and $\pi_G:Q\rightarrow Q/G$ is
a principal bundle. 

We will prove that $J^{-1}(0)/G$ is symplectomorphic to $(T^*(Q/G), \ \omega_{Q/G})$. To
see that, consider the codifferential of the mapping $\pi_G$,
$T\pi_G^*:T^*(Q/G)\rightarrow T^ *Q$ and compose with the natural
projection over the quotient $p:T^*Q\rightarrow T^*Q/G$. Then the
mapping $p\circ T^*\pi_G$ is easily seen to give the desired
identification when restricted to its image, which is
$J^{-1}(0)/G$. Notice that although the codifferential $T^*\pi_G$ is
not a mapping (it is multi-valued), the composition does become an identification.
 
$J^{-1}(\mu)/G_{\mu}$ is known to be symplectomorphic to $(T^*(Q/G), \
\omega_{Q/G}+B_{\mu})$ when $G_\mu=G$ and where $B_\mu$ is a magnetic term. To prove that, take $\alpha_{\mu}$
a $1$-form on $Q$ such that
\begin{enumerate}
\item $\alpha_\mu$ is $G$-invariant by $\Phi^{T^*}$.
\item $J\circ\alpha_\mu=\mu$.
\end{enumerate}
Then we have the shift by $\alpha_\mu$ given by $shift:
T^*Q\rightarrow T^*Q$, such that
$shift(\alpha_q)=\alpha_q-\alpha_\mu$. This mapping is $G$-equivariant
and $shift^*(\omega_{Q/G})=\omega_{Q/G}-d\alpha_\mu$. Moreover, this
map satisfies $shift(J^{-1}(\mu))=J^{-1}(0)$ and thus, by
$G$-equivariance, $\frac{shift}{G}(J^{-1}(\mu)/G)=J^{-1}(0)/G$, where
$\frac{shift}{G}$ is the mapping induced on the quotient. The
right hand side of the last equality is identified with $T^*(Q/G)$ but since $shift$
is not a symplectomorphim between the canonical symplectic structures
of cotangent bundles, the form $\omega_{Q/G}$ must be modified. Since
$J\circ\alpha_\mu=\mu$, then $\alpha_\mu(\xi_Q)=\mu(\xi)$ is a
constant function on $Q$.  We deduce that
\[
i_{\xi_Q}d\alpha_\mu=\mathcal{L}_{\xi_Q}d\alpha_\mu-d(\alpha_\mu(\xi_Q))=0
\]
and so there exists a unique $2$-form on $Q/G$ such that
$\pi_G^*\beta_\mu=d\alpha_\mu$. It is not hard to see now that
$J^{-1}(\mu)/G$ with the symplectic structure provided by the
Marden-Weinstein reduction is symplectomorphic to the cotangent bundle
$T^*(Q/G)$ with the symplectic structure given by
$\omega_{Q/G}+\pi_{Q/G}^*\beta_\mu$.
\begin{remark}{\rm 
In the constructions of our paper, we used a connection from the
beginning. With a connection at hand, the construction of the form
$\alpha_\mu$ is just the composition of the connection $1$-form
(which is a $\mathfrak{g}$-valued $1$-form) and $\mu$.}
\end{remark}
\end{appendices}

\section*{Acknowledgments}
This work has been partially supported by MINECO MTM 2013-42-870-P and the ICMAT Severo Ochoa project SEV-2011-0087.
M. Vaquero wishes to thank MINECO for a FPI-PhD Position. We wish to thank D. Iglesias-Ponte and J.C. Marrero for several useful discussions.

%\nocite{*}
\bibliography{symmetries}
\bibliographystyle{acm}

\end{document}